\documentstyle[12pt]{article}

\def\la{\langle}\def\ra{\rangle}
\def\be{\begin{eqnarray}}\def\bea{\begin{eqnarray}}
\def\ba{\begin{eqnarray}}
\def\ee{\end{eqnarray}}\def\eea{\end{eqnarray}}
\def\ea{\end{eqnarray}}
\def\ben{\begin{eqnarray}}\def\bitem{\begin{itemize}}
\def\een{\end{eqnarray}}\def\eitem{\end{itemize}}
\def\del{\partial}
\def\G0p{$G_0^\prime$}
\def\bi{\bibitem}

\def\N1520{$N^\star (1520)$}
\def\calL{{\cal L}}

\def\prl{Phys. Rev. Lett.}\def\pr{Phys. Rev.}

\def\B#1{{}^{#1}\mbox{B}}

\def\Tr{{\mbox{Tr}}}
\def\del{\partial}

\def\roughly#1{\mathrel{\raise.3ex\hbox{$#1$\kern-.75em%
\lower1ex\hbox{$\sim$}}}}\def\lsim{\roughly<}

\def\B#1{{}^{#1}\mbox{B}}

\def\B0{\mbox{\boldmath $0$}}

\def\bq{\begin{equation}}
\def\eq{\end{equation}}

\def\b3{\mbox{\boldmath $3$}}\def\b6{\mbox{\boldmath $6$}}
\def\roughly#1{\mathrel{\raise.3ex\hbox{$#1$\kern-.75em%
\lower1ex\hbox{$\sim$}}}}\def\lsim{\roughly<}

\renewcommand{\thefootnote}{\fnsymbol{footnote}}
\begin{document}
%\begin{titlepage}
\hfill\today \vskip 2cm
\begin{center}
{\Large\bf The Vector Manifestation and Effective Degrees of
Freedom At Chiral Restoration}\footnote{Based on talks given at
the YITP-RCNP Workshop on ``{Chiral Restoration in Nuclear
Medium}" at Yukawa Institute of Theoretical Physics, Kyoto, Japan,
7-9 October 2002 and the 2002 International Workshop ``Strong
Coupling Gauge Theories and Effective Field Theories (SCGT02)" at
Nagoya University, Nagoya, Japan, 10-13 December 2002.}

  \vskip 1.5cm
   {{\large Mannque Rho}}
 \vskip 0.5cm

 {\it  Service de Physique Th\'eorique, CEA/DSM/SPhT,
Unit\'e de recherche associ\'ee au CNRS, CEA/Saclay,  91191
Gif-sur-Yvette c\'edex, France}

{\it and}

{\it School of Physics, Korea Institute for Advanced Study, Seoul
130-012, Korea}

\end{center}

\vskip 0.5cm

%\centerline{(\today)}
 \vskip 1.5cm

\centerline{\bf Abstract}
 \vskip 1cm
The role of effective degrees of freedom on the vector and
axial-vector susceptibilities and the pion velocity at chiral
restoration is analyzed. We consider two possible scenarios, one
in which pions are considered to be the only low-lying degrees of
freedom -- that we shall refer to as ``standard" -- and the other
in which pions, vector mesons and constituent quarks (or
quasiquarks in short) are the relevant low-lying degrees of
freedom -- that we shall refer to as ``vector manifestation (VM)."
We show at one-loop order in chiral perturbation theory with
hidden local symmetry Lagrangian that while in the standard
scenario, the pion velocity vanishes at the chiral transition, it
instead approaches unity in the VM scenario. If the VM is realized
in nature, the chiral phase structure of hadronic matter can be
much richer than that in the standard one and the phase transition
will be a smooth crossover: Sharp vector and scalar excitations
are expected in the vicinity of the critical point. Some indirect
indications that lend support to the VM scenario, and in
consequence to BR scaling, are discussed.
%\end{titlepage}
%\newpage
%\tableofcontents
\newpage
\renewcommand{\thefootnote}{\arabic{footnote}}
\setcounter{footnote}{0}

\section{Introduction}
In describing the chiral restoration transition at the critical
temperature $T_C$ and/or critical density $n_C$, one of the
essential ingredients is the relevant degrees of freedom that
enter in the vicinity of the critical point. Depending upon what
enters there, certain aspects of the phase transition scenario can
be drastically different. These different scenarios will
eventually be sorted out by experiments or by QCD simulations on
lattice.

The standard way of looking at the problem at high temperature and
low density currently accepted by the majority of the community as
the ``standard picture" is to assume that near the critical point,
the only relevant low-lying degrees of freedom are the
(psudo-)Goldstone pions and a scalar meson that in the $SU(2)$
flavor case, makes up the fourth component of the chiral
four-vector of $SU(2)_L\times SU(2)_R$. For the two-flavor case,
one then maps QCD to an $O(4)$ universality class etc. Here
lattice measurements will eventually map out the phase structure.
On the contrary, at zero temperature and high density, the
situation is totally unclear. In fact as density increases, the
possibility is that one may not be able to talk about
quasipartcles of any statistics at all: The concept of a hadron
may even break down. Unfortunately lattice cannot help here, at
least for now, because of the notorious sign problem.

The situation is markedly different if the vector manifestation
\`a la Harada and Yamawaki~\cite{HY:VM,HY:PR} scenario is viable.
In this picture, certain hadrons other than pions can play a
crucial role in both high temperature and high density with a
drastically different phase structure. In particular, light-quark
vector mesons and constituent quarks (called quasiquarks from here
on) can become the relevant degrees of freedom near the phase
transition, becoming ``sharper" quasiparticles even near the
critical density, thereby increasing the number of degrees of
freedom that enter from the hadronic sector, with -- in sharp
contrast to the standard picture -- narrow-width excitations near
the critical point. This can then lead to a form of phase change
that is a lot smoother than that of the standard scenario.

The objective of my talk is to describe how this can happen.

\section{The vector manifestation}
To approach the chiral symmetry restored phase ``bottom up" from
the broken phase, we need an effective field theory (EFT) that
represents as closely as possible the fundamental theory of strong
interactions, QCD. In fact, according to Weinberg's unproven
theorem~\cite{wein-theorem},  QCD at low energy/momentum can be
encapsulated in an effective field theory with a suitable set of
colorless fields subject to the symmetries and invariance required
by QCD. So the question is how to construct an EFT that captures
the essence of QCD in describing the relevant physics at the phase
transition. In order to do this, we first need to identify the
scale at which we want to define our EFT and the relevant degrees
of freedom and symmetries that we want to implement. I will
consider two possibilities. One is the standard scenario based on
linear sigma model that assumes that the only low-excitation
degrees of freedom relevant to chiral restoration, apart from the
nucleons, are the pions and possibly a scalar (denoted $\sigma$ in
linear sigma model) with all other degrees of freedom integrated
out. The other is the vector manifestation scenario based on
hidden local symmetry (HLS) in which light-quark vector mesons
figure crucially.

In this talk, we are principally interested in what new physics
can be learned in the second scenario which seems to be currently
ignored by the community in the field. We consider for
definiteness the three-flavor case, i.e., $SU(3)_L\times
SU(3)_R$~\footnote{The two-flavor case is a bit more subtle and
has not yet been fully worked out. I will leave that issue for
later.}. In going to nuclear matter and beyond in this scenario,
we must keep vector-meson degrees of freedom $explicit$. This is
because of the vector manifestation (VM)~\cite{HY:VM,HY:PR} for
which vector degrees of freedom are indispensable. To understand
the VM, we consider the HLS Lagrangian~\cite{bandoetal} in which
the pion and vector mesons are the effective degrees of
freedom~\footnote{HLS is essential for the VM since local gauge
symmetry is required for doing a consistent chiral perturbation
calculation in the presence of vector mesons. Other theories where
local gauge symmetry is absent are moot on this issue. See
\cite{HY:PR} for a clear discussion on this point.}. For the
moment, we ignore fermions and heavier excitations as e.g., $a_1$,
glueballs etc. To make the discussion transparent, we consider
three massless flavors, that is, in the chiral
limit~\footnote{Masses and symmetry breaking can be introduced
with attendant complications. Up to date, the effects of quark
masses have not been investigated in detail in this formalism. It
is possible that the detail structure of the phase diagram can be
substantially different from the chiral limit picture we are
addressing here.}. The relevant fields are the (L,R)-handed chiral
fields $\xi_{L,R}=e^{i\sigma/F_\sigma} e^{\mp i\pi/F_\pi}$ where
$\pi$ is the pseudoscalar Goldstone boson field and $\sigma$ the
Goldstone scalar field absorbed into the HLS vector field
$\rho_\mu$, coupled gauge invariantly with the gauge coupling
constant $g$. If one matches this theory to QCD at a scale
${\Lambda_M}$ below the mass of the heavy mesons that are
integrated out but above the vector ($\rho$) meson mass, it comes
out -- when the quark condensate $\la \bar{q}q\ra$ vanishes as in
the case of chiral restoration in the chiral limit -- that
 \be
g(\bar{\Lambda})\rightarrow 0,\ \ a(\bar{\Lambda})\equiv
F_\sigma/F_\pi \rightarrow 1.
 \ee
Now the renormalization group analysis shows that $g=0$ and $a=1$
is the fixed point of the HLS theory and hence at the chiral
transition, one approaches what is called the ``vector
manifestation" fixed point. The important point to note here is
that {\it this fixed point is approached regardless of whether the
chiral restoration is driven by temperature $T$~\cite{HS:T} or
density $n$~\cite{HKR} or a large number of flavors~\cite{HY:NF}}.
At the VM, the vector meson mass must go to zero in proportion to
$g$, the transverse vectors decouple and the longitudinal
components of the vectors join in a degenerate multiplet with the
pions.
\section{Vector and Axial-Vector Susceptibilities}
As an illustration of what new features are encoded in the HLS/VM
scenario, we first consider approaching the critical point in heat
bath. We will come to the density problem later. Specifically
consider the vector and axial vector susceptibilities defined in
terms of Euclidean QCD current correlators as
 \be
\delta^{ab}\chi_V&=& \int^{1/T}_0 d\tau\int d^3\vec{x}\la V_0^a
(\tau, \vec{x}) V_0^b (0,\vec{0})\ra_\beta,\\
\delta^{ab}\chi_A&=& \int^{1/T}_0 d\tau\int d^3\vec{x}\la A_0^a
(\tau, \vec{x}) A_0^b (0,\vec{0})\ra_\beta
 \ee
where $\la \ra_\beta$ denotes thermal average and
 \be
V_0^a\equiv \bar{\psi}\gamma^0\frac{\tau^a}{2}\psi, \ \
A_0^a\equiv \bar{\psi}\gamma^0\gamma^5\frac{\tau^a}{2}\psi
 \ee
with the quark field $\psi$ and the $\tau^a$ Pauli matrix the
generator of the flavor $SU(2)$.
\subsection{\it The standard (linear sigma model) scenario}
The standard picture with pions figuring as the only degrees of
freedom in heat bath has been worked out by Son and
Stephanov~\cite{son}. The reasoning and the result are both very
simple and elegant. They go as follows.

If the pions are the only relevant degrees of freedom near the
chiral transition, then the axial susceptibility (ASUS) for the
system is encoded in the chiral Lagrangian of the form
 \be
\calL_{eff}=\frac{{f_\pi^t}^2}{4}\left(\Tr\nabla_0 U\nabla_0
U^\dagger - v_\pi^2\Tr\del_i U\del_i U^\dagger\right) -\frac 12
\la\bar{\psi}\psi\ra {\rm Re} M^\dagger
U\label{LA}+\cdots\label{Leff}
 \ee
where $v_\pi$ is the pion velocity, $M$ is the mass matrix
introduced as an external field, $U$ is the chiral field and the
covariant derivative $\nabla_0 U$ is given by $\nabla_0 U=\del_0 U
-\frac i2 \mu_A (\tau_3 U +U\tau_3)$ with $\mu_A$ the axial
isospin chemical potential. The ellipsis stands for higher order
terms in spatial derivatives and covariant derivatives. Now given
(\ref{Leff}) as the $full$ effective Lagrangian which would be
valid if it could be given in a local form as is, then the ASUS
would take the simple form
 \be
\chi_A=-\frac{\del^2}{\del\mu_A^2}\calL_{eff}|_{\mu_A=0}={f_\pi^t}^2.
\label{chia}
 \ee
Within the scheme, this is the $entire$ story: There are no other
terms that contribute. That the ASUS is given solely by the square
of the temporal component of the pion decay constant follows from
the fact that the Goldstone bosons are the only relevant degrees
of freedom in the system, with those degrees of freedom integrated
out being totally unimportant. The effective theory of course
cannot tell us what $f_\pi^t$ is. However one can get it from
lattice QCD. To do this, we exploit that at chiral restoration,
the vector correlator and the axial correlator must be equal to
each other, which means that
 \be
\chi_A|_{T=T_c}=\chi_V|_{T=T_c}.
 \ee
Now from the lattice data of Gottlieb et al~\cite{gottlieb}, we
learn that
 \be
\chi_V|_{T=T_c}\neq 0
 \ee
and hence from (\ref{chia}) that
 \be
f_\pi^t\neq 0.
 \ee
Next, we know that the space component of the pion decay constant
$f_\pi^s$ must go to zero at chiral restoration. This is because
it should be related directly to the quark condensate
$\la\bar{q}q\ra$, i.e., the order parameter of the chiral symmetry
of QCD. Thus one is led to the conclusion that at $T=T_c$ the
velocity of the pion must be zero,
 \be
v_\pi\propto f_\pi^s/f_\pi^t\rightarrow 0 \ \ {\rm as} \ \
T\rightarrow T_c.
 \ee

That the pion velocity is zero at the critical point is both
elegant and reasonable. It is analogous to the sound velocity in
condensed matter physics which is known to go to zero on the
critical surface. But the trouble is that there is a caveat here
which throws doubt on the simple result. One might naively think
that the vector susceptibility (VSUS) could also be described by
the same local effective Lagrangian but with the covariant
derivative now defined with the vector isospin chemical potential
$\mu_V$ as $\nabla_0 U=\del_0 U-\frac 12 \mu_V (\tau_3
U-U\tau_3)$.  If the local form of the $effective$ Lagrangian
(\ref{Leff}) is valid as well for the VSUS, one can do the same
calculation as for $\chi_A$, i.e.,
 \be
\chi_V=-\frac{\del^2}{\del\mu_V^2}\calL_{eff}|_{\mu_V=0}.
\label{chiv}
 \ee
Now a simple calculation shows that $\chi_V=0$ for {\it all T}.
This is at variance with the lattice result. It is also
unacceptable on general grounds. Thus either the local effective
Lagrangian is grossly inadequate for the VSUS or else the
assumption that the pions are the only relevant degrees of freedom
is incorrect. Indeed, Son and Stephanov suggest that diffusive
modes in hydrodynamic language that are not describable by a local
Lagrangian can be responsible for the non-vanishing VSUS.
\subsection{\it The HLS/VM scenario}
The situation is dramatically different in the VM
scenario~\cite{HY:VM,HY:PR}. Both the hidden gauge fields and
quasiquark fields enter importantly at the phase
transition~\cite{HKRS}. The reason for this is that the masses of
both tend to zero near the VM fixed point and hence they must
enter on the same footing as the Goldstone pions.

In the HLS/VM scheme, the parameters of the effective Lagrangian
are defined at the matching scale $\Lambda_M$ in terms of the QCD
parameters that encode the vacuum change in heat bath and/or dense
medium. In computing physical observables like the current
correlators, one takes into account both quantum loop effects that
represent how the parameters run as the scale is changed from the
matching scale to the physical (on-shell) scale $and$ thermal
and/or dense loop effects induced in the renormalization-group
flow. Now the former implies an $intrinsic$ temperature and/or
density dependence in the parameters~\footnote{This dependence is
missing in most of the effective field theory calculations that
are based on effective Lagrangians determined in the matter-free
and zero temperature vacuum. Most of the treatments found in the
literature nowadays belong to this category.} -- called
``parametric dependence." The one-loop calculations in \cite{HKRS}
show that both effects are governed by the VM fixed point at
$T\rightarrow T_c$,
 \be
g\rightarrow 0, \ \ a\rightarrow 1.
 \ee
The results~\cite{HKRS} that follow from this consideration are
 \be
f_\pi^t|_{T=T_c}=f_\pi^s|_{T=T_c}=0, \ \
v_\pi|_{T=T_c}=1\label{main1}
  \ee
and
 \be
\chi_A|_{T=T_c}=\chi_V|_{T=T_c}= 2N_f \left[\frac{N_f}{12} T_c^2
+\frac{N_c}{6} T_c^2\right] \ .\label{main2}
 \ee
One can understand these results as follows. As $T\rightarrow
T_c$, both $f_\pi ^{t,s}$ approach $\bar{f}_\pi\propto
\la\bar{q}q\ra$ which approaches zero~\cite{HS:T}. At one loop,
they approach the latter in such a way that the ratio goes to 1,
thereby making the pion velocity approach the velocity of light.
Both $\chi_{V,A}$ get contributions from the flavor gauge vector
mesons and the quasiquarks whose masses approach zero and chiral
symmetry forces them to become equal to each other. Since both the
space and time components of the pion decay constant are
vanishing, they do not figure in the formulas for the
susceptibilities (\ref{main2}).

If realized, the VM scenario will present an interesting phase
structure. It will give a phase diagram drastically different from
the standard sigma model one. For instance, it would imply that
there are a lot more degrees of freedom than the standard picture
just below the critical temperature. We do not know how fast the
masses actually drop as one approaches, bottom-up, the critical
point but if the presently available lattice results are taken at
their face value, then they do not seem to drop appreciably up to
near $T_c$. But it is still possible that they drop to zero in a
narrow window near the critical point in a way consistent with the
VM and account for the rapid increase of energy density observed
in the lattice calculations. In any event, that would provide a
natural explanation of a smooth transition with a possible
coexistence of excitations of various quantum numbers below and
above $T_c$ as seems to be indicated by the MEM analysis of
\cite{hatsuda}.

I must mention that there is a caveat here also. The results
(\ref{main1}) and (\ref{main2}) are one-loop results and one may
wonder whether two-loop or higher orders would not change the
qualitative features. The space component of the pion decay
constant is undoubtedly connected to the chiral order parameter,
so remains zero to all orders but there is no general argument to
suggest that the time part cannot receive non-vanishing
contributions at higher orders. If it did, then we would fall back
to the Son-Stephanov result of a vanishing pion velocity.
\section{Multiplet structure}
The multiplet structure of hadrons implied by the VM at the phase
transition at $T=T_c$ or $n=n_c$ or $N_f=N_f^c$ is basically
different from that of the standard one based on linear sigma
model. Continuing with three flavors in the chiral limit, at the
phase transition where the VM is realized, the longitudinal
components of the vector mesons $\rho_\parallel$ join the
$(1,8)\oplus (8,1)$ multiplet of the Goldstone pions and the
transverse vectors, massless, decouple from the system as the
gauge coupling vanishes. This contrasts with the linear sigma
model picture where a scalar joins the Goldstone pions in
$(3,3^*)\otimes (3^*,3)$ multiplets. If one were to explicitly
incorporate the $a_1$ vector mesons and the scalar meson in the
scheme (so far integrated out), then the $a_1$ would be in the
same multiplet $(3,3^*)\otimes (3^*,3)$ with the scalar in the VM
while in the standard picture, the $a_1$ would be in the multiplet
$(1,8)\oplus (8,1)$ with the vectors $\rho$.

As stressed by Harada and Yamawaki, physically there is no sense
to {\it be on} the VM point. It makes sense only to $approach$ it
from below. One can however ask what happens precisely at the VM.
Since the VM is in the Wigner mode, one may wonder whether the
decoupled vector mesons should not have chiral partners. If they
should, then the HLS/VM would be in difficulty since the theory
does not contain such chiral partners in the same multplet. The
only way that I see to avoid this obstruction is that the
transverse vectors become ``singlet" under chiral transformation.
This can be made possible if the chiral transition is considered
as the flavor vector mesons getting de-Higgsed to the color gauge
bosons, i.e., gluons, as proposed by Wetterich~\cite{wetterich}.
Specifically one can think of the flavor vector mesons as the
vectors excited in the color-flavor locking (CFL) transition
 \be
SU(3)_L\times SU(3)_R\times SU(3)_c\rightarrow SU(3)_{L+R+c}
 \ee
in analogy to the CFL in color superconductivity in QCD at
asymptotic density. In this case, the flavor vectors un-lock the
color and flavor and turn in a sort of relay~\cite{MR:Taiwan} into
the color gauge vector mesons. This phenomenon can be summarized
by writing
 \be
\xi_{L(R)i}^\alpha=[\xi_{L(R)}v]_i^a, \ \ U=\xi_L^\dagger\xi_R
 \ee
in terms of a color-singlet $\xi$ field and a $v\in SU(3)_C$, both
of which are unitary. One can then
relate~\cite{wetterich,MR:Taiwan} the vector meson fields
$\rho_\mu$ and the baryon fields $B$, to the gluon fields $A_\mu$
and the quark fields $\psi$ as
 \be
B&=&Z_\psi^{1/2}\xi^\dagger \psi v^\dagger,\nonumber\\
\rho_\mu&=&v (A_\mu  +\frac{i}{g} \del_\mu)
v^\dagger\label{dressed}
 \ee
where $Z_\psi$ is the quark wave function renormalization
constant. This CF unlocking scenario appears to be highly
appealing and fascinating. Up to date, however, this idea has not
been fully worked out -- the mechanism for CFL and CF-unlocking is
not known -- and although quite plausible, it is not proven yet
that it is not inconsistent with the known structure of QCD. It
remains to be investigated.
\section{Evidences for the VM?}
The VM prediction is clean and unambiguous for $SU(3)_L\times
SU(3)_R$ chiral symmetry in the chiral limit at the chiral
restoration point. At present, there are no lattice measurements
that would validate or invalidate this picture. Are there
experimental indications that Nature exploits this scheme?

So far, there are no indications from relativistic heavy-ion
processes as to whether the VM is realized in the vicinity of the
critical temperature or density. So to answer this question, one
would have to work out what happens at temperatures and/or
densities away from the critical point. One would also have to
consider two-flavor cases and quark mass terms to make contact
with Nature. To do all these is a difficult task and no
theoretical work has been done up to date on this matter.  What is
available up to date are some indications in nuclear systems at
low temperature. Nuclei involve many nucleons in the vicinity of
nuclear matter density and the density regime involved here is
rather far from the density relevant to the VM. Thus we are
compelled to invoke a certain number of extrapolations toward the
VM point. Near nuclear matter density, however, we have a
many-body fixed point known as ``Fermi-liquid fixed
point"~\cite{FR,shankar,MR:Taiwan} which involves quantum critical
phenomenon and this makes the connection to the VM tenuous even
when one is at a much higher density.

To make progress in this circumstance, we have to make a rather
drastic simplification of the phase structure. Here we will adopt
what is called ``double-decimation approximation"~\cite{BR-DD}
which consists of (1) extrapolating downwards from the VM to the
Fermi-liquid fixed point and (2) extrapolating upwards from the
zero-density regime where low-energy theorems apply to the
Fermi-liquid fixed point at nuclear matter density. The spirit
here is close to BR scaling proposed in 91~\cite{BR91}.

Close to the VM, the vector meson mass must go to zero in
proportion to $g$.  Specifically
 \be
m_\rho^*/m_\rho \approx g^*/g \approx
\la\bar{q}q\ra^*/\la\bar{q}q\ra \rightarrow 0
 \ee
as the transition point $n=n_c$ is reached. One can understand
this as follows. Near the critical point the ``intrinsic term"
$\sim g^* F_\pi^*$ in the vector mass formula drops to zero faster
than the dense loop term that goes as $\sim g^* H(n)$ where $H$ is
a slowly (i.e., logarithmically) varying function of density. So
the dense loop term controls the scaling. Now it seems to be a
reasonable thing to assume that $near$ the VM fixed point, we have
the scaling
 \be m_\rho^*/m_\rho \approx g^*/g\approx
\la\bar{q}q\ra^*/\la\bar{q}q\ra.
 \ee
Our conjecture~\cite{BR-DD} is that this holds down to near
nuclear matter density.

Let us now turn to the low-density regime, that is, a density
below nuclear matter density. At near zero density, one can apply
chiral perturbation theory with a zero-density HLS Lagrangian
matched to QCD at a scale $\Lambda_M\sim \Lambda_\chi$. We expect
to have~\cite{BR:yamagishi}
 \be
m_\rho^*/m_\rho \approx f_\pi^*/f_\pi \approx
\sqrt{\la\bar{q}q\ra^*/\la\bar{q}q\ra}.\label{dd1}
 \ee
This result follows from an in-medium GMOR relation for the pion
if one assumes that at low density the pion mass does not scale
(as indicated experimentally~\cite{yamazaki}), that the vector
meson mass is dominantly given by the ``intrinsic term" $\sqrt{a}
F_\pi g$ with small loop corrections that can be ignored and that
the gauge coupling constant does not get modified at low density
(as indicated by chiral models and also empirically). The
double-decimation approximation is to simply assume that this
relation holds from zero density up to nuclear matter
density~\cite{BR-DD}. Note that we are essentially summarizing the
phase structure up to chiral restoration by two fixed points,
namely, the Fermi-liquid fixed point and the vector-manifestation
fixed point. Here we are ignoring the possibility that there can
be other phase changes, such as kaon condensation (or hyperon
matter), color superconductivity etc. which can destroy the Fermi
liquid structure before chiral symmetry is restored.

The one important feature that distinguishes the HLS/VM theory
from other EFTs is the $parametric$ dependence on the background
of the ``vacuum" -- density and/or temperature -- which
intricately controls the fixed point structure of the VM. At low
density, this dependence is relatively weak, so hard to pinpoint.
But in precision experiments, it should be visible. One such case
is the recent experiment of deeply bound pionic atoms. For this,
we can consider a chiral Lagrangian in which only the nucleon and
pion fields are kept explicit with the vectors and other heavy
hadron degrees of freedom integrated out from the HLS/VM
Lagrangian. The relevant parameters of the Lagrangian are the
``bare" nucleon mass, the ``bare" pion mass, the ``bare" pion
decay constant, the ``bare" axial-vector coupling and so on which
depend non-trivially on the scale $\Lambda_M$ and density $n$.
This Lagrangian takes the same form as the familiar one apart from
the $intrinsic$ dependence of the parameters on $n$. (In the usual
approach, the scale $\Lambda_M$ is fixed at the chiral scale and
the dependence on $n$ is absent). As shown by Harada and
Yamawaki~\cite{HY:matching,HY:PR}, the local gauge symmetry of HLS
Lagrangian enables one to do a systematic chiral perturbation
theory even when massive vectors are present. Since the vectors
are integrated out, the power counting will be the same as in the
conventional approach. Now if the density involved in the system
is low enough, say, no greater than nuclear matter density, then
one could work to leading order in chiral expansion. Suppose that
one does this to the (generalized) tree order. To this order, the
parameters of the Lagrangian can be identified with physical
quantities. For instance, the bare pion decay constant $F_\pi$ can
be identified with the physical constant $f_\pi$, the parametric
pion mass with the physical pion mass $m_\pi$ etc. Now in the
framework at hand, the only dependence in the constants on density
will then be the $intrinsic$ one determined by the matching to QCD
immersed in the background of density $n$.

If we apply the above argument to the recent measurement by Suzuki
et al~\cite{yamazaki} of deeply bound pionic atom systems, we will
find that the measurement supplies information on the ratio
$f_\pi^*/f_\pi$ at a density $n\lsim n_0$. There is a simple
prediction for this quantity~\cite{FR,BR:PR01}. We have from
(\ref{dd1})
 \be
\Phi (n)\equiv f_\pi^*/f_\pi \approx
\sqrt{\la\bar{q}q\ra^*/\la\bar{q}q\ra} .\label{dd2}
 \ee
Instead of calculating the quark condensate in medium which is a
theoretical construct, we can extract the in-medium pion decay
constant by extracting $\Phi$ from experiments. Indeed, the
scaling $\Phi$ has been obtained from nuclear gyromagnetic ratio
in \cite{FR,BR:PR01}. At nuclear matter density, it comes out to
be
 \be
\Phi (n_0)=0.78
 \ee with an uncertainty of $\sim 10\%$. Thus it is predicted that
 \be
(f_\pi^*(n_0)/f_\pi)_{th}^2 \approx 0.61.\label{fpith}
 \ee
This agrees with the value extracted from the pionic atom data of
\cite{yamazaki},
 \be
(f_\pi^*(n_0)/f_\pi)_{exp}^2=0.65\pm 0.05.
 \ee

It is perhaps important to stress that this ``agreement" cannot be
taken as an evidence for ``partial chiral restoration" as one
often sees stated in the literature. If one were to go to higher
orders in chiral expansion, the $parametric$ pion decay constant
cannot be directly identified with the physical pion decay
constant since the latter should contain two important
corrections, i.e., quantum corrections governed by the
renormalization group equation as the scale is lowered from
$\Lambda_M$ to the physical scale and dense loop corrections
generated by the flow. At chiral restoration, it is this latter
that signals the phase transition: The parametric pion decay
constant with the scale fixed at the matching scale does not go to
zero even at the chiral restoration point~\cite{HY:PR}. Thus when
one does a higher-order chiral perturbation calculation of the
same quantity, one has to be careful which quantity one is dealing
with.

What one can say with some confidence is that (\ref{fpith}) goes
in the right direction in the context of BR scaling.

A variety of other evidences that lend, albeit indirect, support
to the scaling~\cite{BR91} and in consequence to the notion of the
VM are discussed in \cite{BR:PR01,BR-DD}. If the VM were verified
by going near the chiral transition point, it would constitute a
nice illustration of how the mass of the hadrons making up the
bulk of ordinary matter around us is made to ``disappear," a deep
issue in physics~\cite{wilczek}.
\subsection*{Acknowledgments}

I have benefitted from numerous discussions with Gerry Brown,
Masayasu Harada, Youngman Kim and Koichi Yamawaki.

\end{document}